\documentclass[journal]{IEEEtran}
\usepackage{amsthm}
\usepackage{graphicx}
\usepackage{subcaption}
\usepackage{bbm}
\usepackage{booktabs}
\usepackage{xcolor}

\newtheorem{theorem}{Theorem}
\newtheorem{lemma}{Lemma}

\usepackage{cite}

\usepackage{bm}

\usepackage{amsmath}
\usepackage{amssymb}
\usepackage{mathtools}

\DeclareMathOperator*{\argmax}{arg\,max}
\usepackage{algorithm}
\usepackage{algorithmic}
\makeatletter
\def\BState{\State\hskip-\ALG@thistlm}
\makeatother

\usepackage{setspace}

\usepackage{fixltx2e}

\usepackage{stfloats}

\usepackage{amsthm}
\usepackage{xcolor}
\usepackage{lipsum}
\usepackage[margin=0.75in]{geometry}
\newgeometry{top=1in, left=0.75in,right=0.75in,bottom=0.75in}

\begin{document}
\setlength{\belowcaptionskip}{-10pt}
\setlength{\abovedisplayskip}{2pt}
\setlength{\belowdisplayskip}{2pt}
\setlength{\parskip}{0pt}
%
\title{An Online Scheduling Algorithm for a Community Energy Storage System}

\author{\IEEEauthorblockN{Nathaniel Tucker\IEEEauthorrefmark{2} and
Mahnoosh Alizadeh\IEEEauthorrefmark{2}\\\vspace{0.5cm}}

\IEEEauthorblockA{\IEEEauthorrefmark{2}Department of Electrical and Computer Engineering, 
University of California, Santa Barbara,
California, 93106, USA\\\vspace{0.5cm}}

}


%



\maketitle


\begin{abstract}
In this paper, we consider a community energy storage (CES) system that is shared by various electricity consumers who want to charge and discharge the CES throughout a given time span.  We study the problem facing the manager of such a CES who must schedule the charging, discharging, and capacity reservations for numerous users. Moreover, we consider the case where requests to charge/discharge the CES arrive in an online fashion and the CES manager must immediately allocate charging power and energy capacity to fulfill the request or reject the request altogether. The objective of the CES manager is to maximize the total value gained by all of the users of the CES while accounting for the operational constraints of the CES. We discuss an algorithm titled \textsc{CommunityEnergyScheduling} that acts as a pricing mechanism based on online primal-dual optimization as a solution to the CES manager's problem. The online algorithm estimates the dual variables (prices) in real-time to allow for requests to be allocated or rejected immediately as they arrive. Furthermore, the proposed method promotes charging and discharging cancellations to reduce the CES's usage at popular times and is able to handle the inherent stochastic nature of the requests to charge/discharge stemming from randomness in users' net load patterns and weather uncertainties. Additionally, we are able to show that the algorithm is able to handle any adversarially chosen request sequence and will always yield total welfare within a factor of $\frac{1}{\alpha}$ of the offline optimal welfare. 
\end{abstract}


%
\IEEEpeerreviewmaketitle

\makeatletter
\def\blfootnote{\xdef\@thefnmark{}\@footnotetext}
\makeatother

\blfootnote{
\indent
This work was supported in part by NSF under Grant
1847096 and in part by the UCSB Institute for Energy Efficiency (IEE).}


\section{Introduction}
\label{section: Intro}
Due to the increasing integration of distributed renewable generation in modern power grids, there is growing interest towards implementing distributed \textit{energy storage} (ES) systems in close proximity to energy consumers \cite{ibrahim2008energy, huggins2010energy, brunet2013energy}. Implementing ES near consumers enables various positive outcomes stemming from increased opportunities in demand-side management, e.g., CO$_2$ emission reduction from peak load shaving, increasing the amount of locally-consumed energy from nearby renewable distributed generation, or electricity cost reduction from shifting electricity purchases to off-peak hours \cite{katsanevakis2017aggregated}. \textcolor{black}{Additionally, the concept of \textit{energy communities} is on the rise. Specifically, these are groups of residential and commercial consumers/prosumers that cooperate and take advantage of shared resources (e.g., energy storage systems \cite{SES_in_ES, ntucker_JLQC}) and make use of each others' excess renewable generation. Recently, energy communities have garnered much research interest in various areas, including, but not limited to, peer-to-peer energy trading \cite{p2p1},\cite{p2p2}, blockchain based energy transactions \cite{blockchain1},\cite{blockchain2}, real-time optimization for energy management \cite{realtime1new}, \cite{ntucker_allerton}, and game-theoretic market designs \cite{gametheoretic1new}.  }

To maximize the utility gained from distributed ES implementations and energy communities, the concept of \textit{community energy storage} (CES) is increasing in popularity \cite{parra2017interdisciplinary, klein2016building}. Specifically, a CES is a modular ES implemented within an energy consumption area (e.g., neighborhood, shopping center, etc.) in combination with renewable distributed generation in the area. CES systems are larger than single-consumer ES systems and have larger technical and economic benefits than single-consumer systems due to diversity in load profiles, removing the need for personal  investments by individual consumers, as well as economies of scale \cite{parra2017interdisciplinary, 8025410, 9239974}.   Recently, there has been much work focusing on optimizing the design \cite{parra2016optimum} and the basic operation \cite{terlouw2019multi, van2018techno} of CES. A comprehensive review of different aspects of modern CES can be found in \cite{dai2021utilization}.

While it is evident that CES has great potential to positively impact energy consumers, the effectiveness of a CES system can be severely limited if it is not operated well. Namely, because there are multiple users who want to take advantage of a CES, there must be a smart management system in place to schedule the users' charging and discharging of the CES. If there is no smart management system in place, the CES might be underutilized or overutilized at various times. For example, all the users of the CES might choose to charge and discharge at similar times, (i.e., charging the CES with excess solar generation midday and discharging in the early evening) which limits the number of users who are able to make use of the CES and potentially leaves the CES underutilized at all other time periods. Additionally, any CES management system also has to deal with large amounts of uncertainty. Currently, one of the major technical challenges for future CES implementations is the requirement to handle uncertainty \cite{lampropoulos2020review} in the charging, discharging, and storage demands of the users. The users of a CES  have inherently stochastic electricity demand and their desired usage of the CES is unknown and time-varying as users' net load patterns can vary significantly from day to day and weather can affect  distributed renewable generation. With this in mind, it is clear that future CES implementations require advanced scheduling algorithms in order to operate effectively (i.e., maximize value gained by the system) under uncertain usage patterns.

\subsection{Main Contributions}
\label{section: main con}

The work presented in this manuscript considers the problem of a CES manager attempting to schedule the charging and discharging of a CES for a group of users. Our proposed solution allows for the users to request temporal charging and discharging profiles from the CES in real-time (as they learn about their needs) and the CES manager is able to immediately accept or deny a request and, if accepted, select the profile that maximizes the users utility. Additionally, due to the fact that our solution handles charging and discharging profiles instead of pure capacity requests, our heuristic is able to promote diverse charging and discharging patterns via dynamically updated prices to exploit charging/discharging cancellations and increase the CES's utilization. For example, a charge/discharge cancellation occurs when user A commits to charging the CES at a given time and user B commits to discharging the CES at the same time, thus effectively cancelling each other's power usage of the CES at that time and allowing other users access to charge/discharge at that time slot. Furthermore, we present a theoretical guarantee on the performance of our heuristic which operates in real-time without knowledge of future requests. We are able to bound the worst case performance of our \textit{online} solution in relation to the \textit{offline} optimal solution (i.e., if the CES manager had known the entire sequence of CES requests beforehand) in the form of a competitive ratio. We note that this is a worst case performance guarantee that holds for any \textit{adversarially} chosen CES request sequence.  

The main contributions of this paper are as follows:
\begin{itemize}
    \item \textit{Temporal User Flexibility:} The proposed online heuristic allows users to submit requests for temporal charging and discharging profiles from the CES in real-time instead of committing to long-term capacity reservations far in advance. Furthermore, the CES scheduling heuristic will immediately accept or deny the request.
    \item \textit{Charging and Discharging Cancellation:} As stated previously, the proposed scheduling heuristic deals with temporal charging and discharging profiles instead of capacity reservations. This allows for the heuristic's dynamically updated prices to promote diverse charging and discharging schedules of the users to take advantage of concurrent charging and discharging requests cancelling each other out, hence increasing efficiency (explained further in Section \ref{subsection: charging and discharging cancellation}).
    \item \textit{Upholding CES Constraints:} The proposed online heuristic makes use of dynamically updated prices that are designed to ensure that the CES constraints (e.g., maximum charging power, maximum discharging power, maximum capacity) are met at all times.
    \item \textit{\textcolor{black}{Unknown Nature of Future Requests}:} The proposed online heuristic readily handles the inherent uncertainty of the CES scheduling problem including unknown request times, unknown charging/capacity requests, and unknown valuations without the need of a future model. \textcolor{black}{Specifically, we develop an online primal-dual optimization framework (an overview of primal-dual approaches for solving large-scale optimization problems can be found in \cite{komodakis2015playing}) that is able to provide a worst-case performance guarantee for any adversarially selected input sequence. The developed online optimization framework is akin to algorithmic posted pricing mechanisms for online combinatorial auctions. }
    \item \textit{Theoretical Worst Case Performance Guarantee:} The online heuristic is robust to adversarially chosen request sequences and always yields social welfare within a factor of $\frac{1}{\alpha}$ of the offline optimal (i.e., if the CES manager had known the entire sequence of CES requests beforehand).
\end{itemize}

The remainder of the paper is organized as follows: Related works are discussed in Section \ref{section: related works}. Section \ref{section: Preliminaries} presents the CES manager's objective as well as the problem formulation. Section \ref{section: online scheduling heuristic} describes the proposed online scheduling heuristic as a solution to the CES manager's problem and presents the full procedure of the scheduling heuristic as well as a theoretical worst case performance guarantee. Section \ref{sec: numerical results} presents two numerical examples showcasing the scheduling heuristic.

\subsection{Related Works}
\label{section: related works}

A number of recent studies have proposed methodologies for optimizing shared ES at the end-user side. Specifically, \cite{yao2016stochastic} presents a game-theoretic approach to managing a shared ES where users are competing for limited capacity and \cite{yang2021optimal} presents a coalition game formulation for the sizing, operation, and cost allocation of a shared ES with multiple investors. Additionally, \cite{9187984} presents a  Nash bargaining based benefits sharing model for energy cooperation between users and a CES and is focused on the presence of `cheaters' within the system, attempting to gain additional benefits by providing dishonest information. Centralized control of such a shared ES is studied in \cite{yao2015optimal}, but the solution method does not scale with the number of participants and is approximated instead. The authors of \cite{oh2020reinforcement} present a reinforcement learning approach to manage the operation of an ES under uncertain conditions stemming from wind generation. In \cite{walker2021design}, a stochastic optimization is formulated to manage the operation of multiple shared ES systems and the performance of their proposed control policy is compared to the deterministic optimal solution via numerical experiments; however, there is no theoretical performance guarantee (i.e., bounding the gap between the cost of the deterministic optimal solution and the cost of the proposed policy). Papers \cite{tushar2016energy} and \cite{chen2017energy} also study shared ES strategies, and both make use of models that disallow users to increase or decrease their allotted capacity in real-time. Similarly, \cite{liu2017decision} presents a business model for a shared ES that promotes diverse charging/discharging schedules, but the users' capacity reservations are constrained to remain constant across days, thus limiting flexibility. \textcolor{black}{Additionally, \cite{posted_price} studies a posted price mechanism for energy customers arriving in an arbitrary manner and choosing to either purchase a certain amount of energy based on the posted price, or leave without buying. The mechanism has similarities to the one in this manuscript; however,  \cite{posted_price} focuses on  the case of transactive electric vehicle charging rather than scheduling the charging/discharging of a CES.}

There are two papers closest to our work. First, \cite{zhong2020multi} presents a distributed combinatorial auction approach to schedule capacity, charging, and discharging power for a shared ES. In this work, the solution method is allowed to violate the ES's total capacity limit and the over-capacity energy must be purchased from the local grid. Second, \cite{zhao2019virtual} presents a pricing mechanism to sell `virtualized' portions of a shared ES each day. In this work, the prices are selected to be constant for each optimization period, which is simple to implement but limits the ability to promote diverse charging/discharging patterns from the users in real-time. Different from \cite{zhong2020multi} and \cite{zhao2019virtual}, our goal is to present a scheduling heuristic that never violates CES constraints (i.e., does not allocate more capacity than the CES has available and then purchase the over capacity power from the local grid) and makes use of dynamically updated prices that increase and decrease depending on the current utilization of the CES (i.e., dynamically increases prices at times when utilization is high to discourage usage and decreases prices at times when utilization is low to promote usage).

\section{System Model}
\label{section: Preliminaries}

\subsection{CES Manager's Objective}
\label{subsection: objective}

In this section, we describe the problem facing the manager of a community energy storage (CES) system attempting to optimize the energy storage (charging and discharging) schedules for a group of diverse users. Specifically, the objective of the CES manager is to maximize the total value gained by all of the users by optimizing the usage of the system and incentivizing diverse user schedules in order to maximize the benefits delivered by the capacity-limited CES.

In the following, we consider a singular CES that is co-located with potential users of the system in a neighborhood, shopping center, or business park (we note that this work can be readily extended to account for numerous energy storage systems throughout a given area). We assume that each user of the CES has the physical infrastructure in place to charge and discharge the CES at any time and each user may or may not be equipped with behind-the-meter renewable generation. Additionally, we assume that each user has the ability to communicate with the CES manager to submit requests to charge and discharge. 
\begin{figure}[]
    \centering
    \includegraphics[width=0.9\columnwidth]{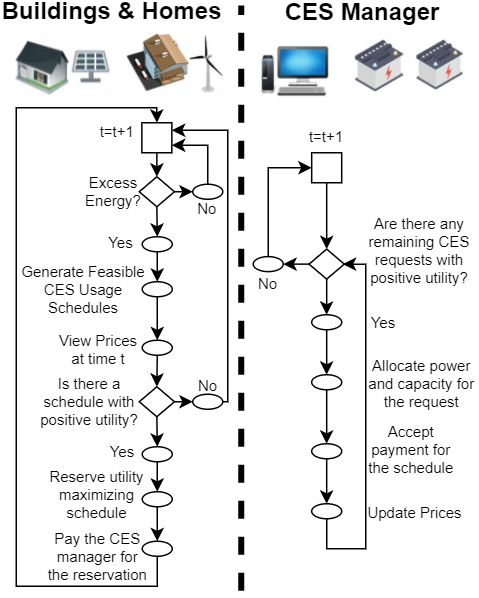}
    \caption{\textcolor{black}{Left: System interactions for CES users (buildings and homes). Right: System interactions for the CES Manager.} }
    \label{fig:system_model}
\end{figure}
\textcolor{black}{A system model can be viewed in Figure \ref{fig:system_model}.}

Over the time span $t=1,\dots,T$, the CES manager receives   $n = 1,\dots, N$ requests to use the shared energy storage system. We note that $N$ is a priori unknown to the CES manager as the CES users are inherently stochastic agents and the CES manager does not know how many requests will be submitted in the time span. In this work, each request $n$ to use the CES is in the form of a temporal charging and discharging profile. Specifically, users submit potential schedules for charging the CES, storing the charged power for a duration, and then discharging the CES at a future time as described in Section \ref{subsection: energy storage schedules}.

The job of the CES manager is to either accept and allocate storage capacity to each request $n$, or to deny the request. Furthermore, due to the stochastic nature of distributed renewable generation and unknown factors affecting users' power consumption, we assume that the users cannot submit their charging/discharging requests far in advance and the CES manager cannot create the usage schedule ahead of time. Rather, users submit charging/discharging requests to the CES manager at random times throughout the time span and the CES manager must make the scheduling decisions immediately, so the users can begin charging and discharging the CES. This means that the allocation algorithm must work in real-time and without knowledge of future requests.

\subsection{Charging and Discharging Schedules}
\label{subsection: energy storage schedules}

Each CES request begins when a user would like to store some energy (e.g., from cheap electricity rates or from excess renewable generation) and make use of it at a later time. A user submitting a storage request may benefit from multiple possible charging/discharging schedules, each providing a different value to the user. The user provides the CES manager with the list of such schedules and the value she associated to each of them. In the end, only a  single schedule may be accepted for the user's charging and discharging of the CES.  For example, if a user has excess solar generation available from 1:00pm-2:00pm and wants to charge the CES, then they could potentially benefit from discharging the power at numerous time periods later in the day, e.g., discharging 5:00pm-6:00pm, 5:30pm-6:30pm, or 6:00pm-7:00pm, etc., each providing different value to the user. Let us define the mathematical notation associated with each request. At time $t_{n}^-$, when the user submits a CES usage request, a set of potential charging and discharging schedules, $\mathcal{S}_n$, is created for request $n$. Each potential charge/discharge schedule $s\in\mathcal{S}_n$ has the following parameters: 

\begin{enumerate}
    \item $t_{n}^-$: The start time for all potential charge/discharge schedules for request $n$.
    
    \item $t_{ns}^+$: The end time for charge/discharge schedule $s$ for request $n$ (Note that the potential charge/discharge schedules need not share an end time). 
    
    \item $i_{nsc}(t)$: The CES charging power profile for request $n$ in feasible schedule $s$ at time $t$. Positive values of $i_{nsc}(t)$ denote that the user is charging the CES and negative values of $i_{nsc}(t)$ denote that the user is discharging from the CES. Note that  $i_{nsc}(t)|_{t=1,\dots,T}$ describes the complete power profile across the entire time span that is to be \textit{charged into} and \textit{discharged from} the CES by user $n$ in feasible schedule $s$.
    
    
    \item $i_{nse}(t)|_{t=1,\dots,T}$: The CES capacity that must be reserved for request $n$ in feasible schedule $s$ across the time span in order to serve the schedule's charging profile.
    
    \item $v_{ns}$: The value of potential schedule $s$ to the user who submitted request $n$. This value is described in detail at the end of this subsection.
\end{enumerate}

\noindent \textit{Example:} Consider the following simple example with a user submitting a request to charge the CES in the time period 8:00am-9:00am at 5kW and then discharge from the CES in the time period 10:00am-11:00am at 5kW and values this schedule at $\$0.50$. Furthermore, let us assume that $t=0$ corresponds to 8:00am, $t=1$ corresponds to 9:00am, $t=2$ corresponds to 10:00am, and $t=3$ corresponds to 11:00am (Note that 1 hour intervals are for simplicity of the example and an implementation would use smaller intervals, 1min, 5min, etc.). As such, the requested schedule's parameters are as follows: 
\begin{enumerate}
    \item Start time: $t_n^- = 0$
    \item End time: $t_n^+ = 3$
    \item Charging profile (kW):  $i_{nc}(t)|_{t=0,1,2,3} = 5,0,-5,0$
    \item CES capacity profile (kWh):  $i_{ne}(t)|_{t=0,1,2,3} = 5,5,5,0$
    \item User valuation (\$): $v_n = 0.50$
\end{enumerate}

As we will see, by exploiting of the CES capacity profile $i_{nse}(t)|_{t=1,\dots,T}$ and the charging power profile $i_{nsc}(t)|_{t=1,\dots,T}$, our algorithm allows the CES manager  to optimize the usage of the CES to avoid overutilization at popular times, underutilization at unpopular times, and to incentivize diverse charge/discharge patterns such that users' requests cancel one another. \textcolor{black}{Additionally, we note that there is no restriction on how many requests per day that a user can submit. If a user submits a request at 8:00am, they can submit multiple other different requests later in the day that would be independent of their earlier requests.}

Before we move on, let us discuss how the valuations $v_{ns}$ can potentially be assigned by the users. As stated previously, there are various potential strategies for energy consumers to make use of CES. For example, users can employ the CES to shift their electricity purchases to take advantage of inexpensive electricity rates during off-peak hours, or users can employ the CES to store locally generated renewable energy and use it at a later time. In all cases, in order for a user to choose to make use of the CES instead of defaulting to purchasing electricity from the grid, there must be an incentive to do so. In this work, we assume that the users are incentivized via cost savings; specifically, a user will only request a charge/discharge profile from the CES if the total cost that the user must pay to the CES manager is less than the cost of purchasing the same energy from the grid. The value  $v_{ns}$ is equal to the the magnitude of such cost savings as discussed next.

The proposed CES scheduling heuristic requires the users' submissions of their valuations of each potential CES schedule (potential charge/discharge schedule parameter 5 listed previously). For the purposes of this paper, we assume the users' motivation to use the CES is to store excess solar energy (that was generated on-site at no cost) to use during later time periods or to charge the CES using inexpensive grid energy and discharge from the CES later in the day to avoid expensive electricity rates. For on-site solar usage, a user's valuation of potential schedule $s$ is equivalent to the cost of electricity from the grid that is replaced by the stored solar:
\begin{equation}
\label{eqn: valuation}
    v_{ns} = -\sum_t p_{\textrm{grid}}(t) i_{nsc}(t)|_{i_{nsc}(t)<0}
\end{equation}
where $p_{\textrm{grid}}(t)$ is the price of electricity from the grid at time $t$ and the negative values of $i_{nsc}(t)$ are the discharging power from the CES. If the user wanted to charge the CES during cheap electricity rates and discharge during expensive electricity rates, the valuation (e.g., cost savings) of such a CES schedule would be calculated as:
\begin{align}
\label{eqn: valuation2}
    \nonumber v_{ns} = -\sum_t& p_{\textrm{grid}}(t)i_{nsc}(t)|_{i_{nsc}(t)<0}
    \\ -&\sum_t p_{\textrm{grid}}(t) i_{nsc}(t)|_{i_{nsc}(t)>0}.
\end{align}

Consider the following example where a user would like to store 5 kWh of locally generated solar energy in the CES from 3:00pm to 5:00pm and then discharge 5 kW from 5:00pm to 6:00pm. Furthermore, assume the local grid's electricity rate from 5:00pm to 6:00pm is 0.11 (\$/kWh). As such, in order for the user to prefer the CES instead of purchasing the electricity from the grid, the total cost for utilizing the CES must be less than $5 \textrm{ kW} \times 1 \textrm{ hour} \times 0.11  \textrm{ \$/kWh} = \textrm{ \$ }0.55 $. In other terms, we can say that the user values that specific CES charge/discharge profile at \$0.55. 


\subsection{CES Constraints}
\label{subsection: SESS model}

The community energy storage system has three parameters that constrain its operation\footnote{\textcolor{black}{We note that we do not include a battery model nor degradation in this manuscript; however, this could easily be added to the framework. Any battery model limitations would reduce the number of feasible charging schedules $s \in \mathcal{S}_n$ and degradation costs could be included in the users' payment calculation (i.e., an extra term could be added to the payment $\tilde{p}_{ns^*}$ calculated in line 9 of Algorithm 1 to account for degradation costs).}}: 1) the CES can store up to $\hat{E}$ kWh at any given time, 2) the CES's maximum charging power $\hat{P}_c$ kW, and 3) the CES's maximum discharging power $\hat{P}_d$ kW. At any given time, the total stored energy, total charging power, and total discharging power of all the users combined must be less than the aforementioned parameters.

\subsection{Charging and Discharging Cancellation}
\label{subsection: charging and discharging cancellation}

One important characteristic of the energy storage scheduling problem is that different users' requests to charge and discharge the CES can occur during the same time period, thus resulting in charge/discharge cancellations. As mentioned in Section \ref{section: Intro}, a charge/discharge cancellation occurs when user A commits to charging the CES at a given time and user B commits to discharging the CES at the same time, thus effectively cancelling each other's power usage of the CES at that time and allowing other users access to charge/discharge at that time slot.



The importance of charging and discharging cancellations is twofold. First, the occurrence of a charging and discharging cancellation decreases the total power being charged/discharged from the CES (recall the CES has a maximum power constraint); therefore, allowing other users access to that time slot. Second, a charge/discharge cancellation eliminates the usage of the CES altogether and instead users within the community are providing power to one another directly. That is, locally generated renewable power that would have been injected into the grid or stored in the CES is instead being used immediately by another user within the community. 

\subsection{Offline and Online Problem}
\label{subsection: offline and online}
In the body of this work, we first formulate the energy storage scheduling problem as an \textit{offline} optimization and then use the offline problem to aid the design of a heuristic to solve the \textit{online} problem. In the offline case, we assume the CES manager is clairvoyant and knows the entire sequence of $N$ energy storage requests over the time span $t=1,\dots,T$. As such, the offline CES manager can create the optimal schedules for the energy storage requests and can achieve maximal value. However, the reality is that the CES manager does not know the users' desired charging and discharging times and storage capacity needs in advance. Instead, the energy storage requests are revealed one-by-one throughout the time span meaning that an online solution method is required for real world implementation. Additionally, the energy storage scheduling problem has obstacles that are not easily overcome in many online heuristics; namely, the lack of accurate statistics for the users' energy storage requests as there are many exogenous factors that directly affect the time and capacity of such requests (e.g., stochastic renewable generation and weather affect the time and capacity of energy storage requests and random human behavior affects desired discharging times). As such, in the following we present an online solution that can account for \textit{adversarially} chosen sequences of energy storage requests and still yield utility that is within a constant factor of the clairvoyant offline solution. Let us first state the offline problem.

\subsection{Offline Problem Formulation}
\label{subsection: offline problem formulation}
The state of the CES at any time $t$ can be fully described by the following two variables: $y_e(t)$ the total energy capacity that is reserved at time $t$ summed across all requests and $y_c(t)$ the total charging power that is scheduled for time $t$. In order to calculate $y_e(t)$ and $y_c(t)$, we introduce the decision variable $x_{ns}$. Specifically, when request $n$ to use the CES is received, the CES manager must select one of the potential schedules $s\in\mathcal{S}_n$ or deny the request altogether. As such, the CES manager sets the variable $x_{ns}$ equal to 1 if schedule $s$ is selected for request $n$ and 0 otherwise. If no CES schedule is selected, the request is denied and $x_{ns}=0,\;\forall s$.

The total demands for energy capacity and charging power, $y_e(t)$ and $y_c(t)$ respectively, are calculated as follows:

\begin{align}
    \label{eqn:y_e}
    & y_{e}(t)=\sum_{\mathcal{N},\mathcal{S}_n} i_{nse}(t) x_{ns},\\
    \label{eqn:y_c}
    & y_{c}(t)=\sum_{\mathcal{N},\mathcal{S}_n} i_{nsc}(t) x_{ns}.
\end{align}

As stated in Section \ref{subsection: offline and online}, if the CES manager has full knowledge of the sequence of CES requests, the optimal schedules can be found by solving the following \textit{offline} optimization:

\begin{subequations}
\begin{align}
    &\max_{x} \sum_{\mathcal{N},\mathcal{S}_n} v_{ns}x_{ns}
    \label{eqn:offline obj}
    \\
    &\nonumber\textrm{subject to:}
    \\
    &x_{ns} \in \{0,1\},
    && \forall{n\in\mathcal{N},s\in\mathcal{S}_n}
    \label{eqn:offline integer constraint}
    \\
    &\sum_{\mathcal{S}_n }x_{ns} \leq 1, && \forall{n\in\mathcal{N}}
    \label{eqn:offline 1 sched}
    \\
    &y_e(t) \leq \hat{E},
    && \forall{t\in\mathcal{T}}
    \label{eqn:offline energy capacity constraint}
    \\
    &y_c(t) \leq \hat{P}_c,
    && \forall{t\in\mathcal{T}}
    \label{eqn:offline charging power constraint}
    \\
    &y_c(t) \geq -\hat{P}_d,
    && \forall{t\in\mathcal{T}}.
    \label{eqn:offline discharging power constraint}
\end{align}
\end{subequations}

In \eqref{eqn:offline obj}, the objective is to maximize the total value of CES schedules across all requests. Constraint \eqref{eqn:offline integer constraint} is an integer constraint on the decision variable. Constraint \eqref{eqn:offline 1 sched} ensures that only one CES usage schedule can be selected per request. Constraints \eqref{eqn:offline energy capacity constraint}, \eqref{eqn:offline charging power constraint}, and \eqref{eqn:offline discharging power constraint} enforce the energy capacity limit, charging power limit, and discharging power limit of the CES, respectively.

Furthermore, to gain insight into how to formulate an online pricing heuristic for the CES problem, the offline optimization can be examined in the dual domain\footnote{We note that the integer constraint \eqref{eqn:offline integer constraint} must be temporarily relaxed in order to formulate the offline dual. However, we also note that our competitive ratio results for our online pricing mechanism are for integer allocations.}. Specifically, we make use of Fenchel Duality and use the dual variables $u_n$, $p_e(t)$, $p_c(t)$, and $p_d(t)$ \cite{Fenchel}. \textcolor{black}{The dual variable $u_n$ corresponds to the utility gained by the user who submitted request $n$. That is, their valuation of their assigned energy storage schedule minus the price of that schedule that they pay to the CES manager. We note that each user’s utility should be positive if they are using the CES and 0 if their request is denied. Additionally, the dual variables $p_e(t)$, $p_c(t)$, $p_d(t)$ are associated with the total energy capacity constraint, total charging power constraint, and total discharging power constraint, respectively. Moreover, they can be viewed as the marginal prices that the users must pay for utilizing the limited storage, charging power, and discharging power of the CES.} Additionally, in the remainder of the paper, the Fenchel conjugate of a function $f(y(t))$ is defined as:

\begin{equation}
    f^*(p(t)) = \sup_{y(t)\geq0} \big\{ p(t)y(t) - f(y(t)) \big\}.
\end{equation} 

\textcolor{black}{In this work and many other online combinatorial problems, making use of Fenchel conjugate functions yields a generalized dual problem that can be used to design online solution algorithms (e.g., online packing/covering \cite{buchbinder2009online}, online paging/caching \cite{bansal2008randomized}, online
matching \cite{buchbinder2007online}, etc.). Namely, the conjugate functions $f^*(p(t))$ that appear in the Fenchel dual problem's objective function could account for various convex cost functions due to increasing usage of limited resources\footnote{In this work, we do not explicitly make use of cost functions for utilizing limited resources (capacity and power); however, the capacity and power constraints' costs could be viewed as zero-infinite step functions, which would yield the same Fenchel conjugates as \eqref{eqn: fenchel energy}-\eqref{eqn: fenchel discharge}.} or scaling penalties. Furthermore, we note that the Lagrange dual is a special case of the more general Fenchel dual problem; moreover, the Fenchel dual can be derived from the Lagrange dual problem and the conjugate definition (shown in \cite{huang2014sigact}). We note that Lagrangian duality is used in similar primal-dual works \cite{anand2012resource, gupta2012online, nguyen2013lagrangian}; however, in general, the Fenchel dual typically presents a better structure for the design and analysis of online primal-dual algorithms \textcolor{black}{that attempt to approximate solutions for NP-hard combinatorial problems such as the one we study in this work.} We refer the reader to \cite{komodakis2015playing, Fenchel, devanur2017primal, IaaS, huang2014sigact} for further reading on Fenchel duality in this setting and primal-dual methods.}

With the aforementioned dual variables and Fenchel conjugate definition, the offline Fenchel dual of \eqref{eqn:offline obj}-\eqref{eqn:offline discharging power constraint} is as follows:

\begin{subequations}
\begin{align}
    \label{eqn: dual obj}
    &\min_{u,p} \sum_{\mathcal{N}} u_{n}
    + \sum_{\mathcal{T}}\Big[ f_{e}^*(p_{e}(t))
    +  f_{c}^*(p_{c}(t))
    +  f_{d}^*(p_{d}(t))\Big]
    \\*
    &\nonumber\textrm{ subject to:}
    \\*
    \label{eqn: dual utility const}& u_{n} \geq v_{ns} - \sum_{\mathcal{T}}\Big[i_{nse}(t)p_e(t)
    + i_{nsc}(t)p_c(t)\\
    &\nonumber\hspace{38pt}- i_{nsc}(t)p_d(t)\Big],
     \hspace{17pt}\forall{s\in\mathcal{S}_n, n\in\mathcal{N}}
    \\*
    &\label{eqn: dual utility const nonnegative}u_n \geq 0,
     \hspace{92pt}\forall{n\in\mathcal{N}}
    \\*
    &\label{eqn: dual prices nonneg}p_e(t), p_c(t), p_d(t) \geq 0,
     \hspace{32pt}\forall{t\in\mathcal{T}}.
\end{align}
\end{subequations}
We note that $f_{e}^*(p_{e}(t))$, $f_{c}^*(p_{c}(t))$, and $f_{d}^*(p_{d}(t))$ are the Fenchel conjugates for the energy capacity limit, charging power limit, and discharging power limit, respectively. Recall from Section \ref{subsection: SESS model}, 1) the CES can store up to $\hat{E}$ kWh at any given time, 2) the CES's maximum charging power $\hat{P}_c$ kW, and 3) the CES's maximum discharging power $\hat{P}_d$ kW. With these variables and the dual variables $p_e(t)$, $p_c(t)$, and $p_d(t)$, the Fenchel conjugates can be written as follows:

\begin{align}
\label{eqn: fenchel energy}
    f_e^*(p_e(t)) = \hat{E}\;p_e(t),
\end{align}

\begin{align}
    f_c^*(p_c(t)) = \hat{P}_{c}\;p_c(t),
\end{align}

\begin{align}
\label{eqn: fenchel discharge}
    f_d^*(p_d(t)) = \hat{P}_{d}\;p_d(t).
\end{align}

\subsection{Insight on Scheduling Decisions}
\label{subsection: Scheduling Decisions}

In order to learn how to make scheduling decisions in the online case, let us first examine the offline Fenchel dual \eqref{eqn: dual obj}-\eqref{eqn: dual prices nonneg}. The constraint \eqref{eqn: dual utility const} gives insight into the optimal scheduling decisions for each request $n$. Specifically, if the utility gained $u_n$ from request $n$ is negative across all potential schedules, then the request to utilize the CES is denied \textcolor{black}{and $u_n$ is set equal to 0}. However, when $u_n > 0$ then the request is accepted and the charging/discharging/storage schedule $s\in\mathcal{S}_n$ to be selected is the one that returns the maximal $u_n$. \textcolor{black}{With this in mind, we can instead use the following equation to calculate the utility of request $n$:}

\begin{align}
\label{eqn: u_n}
    \nonumber u_n = \max \bigg\{ &0, \max_{\mathcal{S}_n} \Big\{v_{ns}
    - \sum_{\mathcal{T}} \big[ i_{nse}(t)p_e(t)
    \\& + i_{nsc}(t)p_c(t)
    - i_{nsc}(t)p_d(t) \big] \Big\}\bigg\}.
\end{align}

\textcolor{black}{Equation \eqref{eqn: u_n} is derived from examining the KKT conditions for the dual problem \eqref{eqn: dual obj}-\eqref{eqn: dual prices nonneg}. Specifically, for any request $n$ to use the CES, there is a dual variable $u_n\geq0$ from constraint \eqref{eqn: dual utility const} which corresponds to the utility of request $n$. Moreover, we know that in the offline primal and dual solutions, no schedule can be selected unless constraint \eqref{eqn: dual utility const} is tight for a specific schedule. As such, we can set the utility equal to the maximum of 0 (corresponding to no schedule being selected due to negative utility gain) and the RHS of \eqref{eqn: dual utility const} (corresponding to utility maximizing schedule being selected). In summary, if the dual variables (CES resource prices) $p_e(t)$, $p_c(t)$, and $p_d(t)$ are known or estimated, then equation \eqref{eqn: u_n} can be used to determine which schedule gets allocated for request $n$ or if request $n$ is denied altogether and $u_n$ is set to 0 (we note that in Section \ref{section: online scheduling heuristic} we present our methodology to estimate the dual variables/CES resource prices in real-time so that \eqref{eqn: u_n} can be solved in an online fashion).}

We note that in order to solve for the utility gained $u_n$ from the offline dual \eqref{eqn: dual obj}-\eqref{eqn: dual prices nonneg} (and the offline primal \eqref{eqn:offline obj}-\eqref{eqn:offline discharging power constraint}), this requires full knowledge of the requests to use the CES beforehand. However, as discussed previously in Section \ref{subsection: offline and online}, the manager of the CES does not know the sequence of requests beforehand and must make scheduling decisions as they arrive without knowledge of future requests. \textcolor{black}{Moreover, as we show in the remainder of the paper, we never have to solve the offline dual problem as presented in (7a)-(7d), as this would require knowledge of the entire sequence of usage requests, which the CES manager does not have. Instead, we make use of dual variable update functions (12), (15), (16) to estimate the dual variables in real-time. Then, these dual variables are used as ‘prices’ for the limited resources and our algorithm selects schedules  w.r.t. these prices. We can show that our estimated dual variables will always yield feasible solutions to the primal problem. This is because the dual variable update functions are carefully selected to yield values that increase as the usage of the CES increases. Then, when a constraint is about to be violated, the dual variable update functions will output values high enough such that no energy storage schedule yields positive utility, meaning that requests will be denied if constraints are going to be violated. The gap in the objective value from the original (unrelaxed) primal problem (5a)-(5f) and our online heuristic is bounded in Theorem 1.} 


\section{Online CES Scheduling Heuristic}
\label{section: online scheduling heuristic}

\subsection{Online Scheduling via Dual Variable Updates}
\label{subsection: online scheduling}

In the following, we present a scheduling heuristic for optimizing usage of the CES that updates the dual variables $p_e(t)$, $p_c(t)$, and $p_d(t)$ in an online fashion as requests are revealed. Then, with the estimated dual variables, the algorithm solves equation \eqref{eqn: u_n} for each request to select the utility maximizing charging/discharging/storage schedule. Moreover, the online scheduling heuristic updates the dual variables for charging, discharging, and storage based only on $y_e(t)$ and $y_c(t)$ (the total energy capacity reserved at time $t$ and the total charging power scheduled at time $t$, respectively). 

The online scheduling procedure for the usage of the CES is outlined in Algorithm \textsc{CommunityEnergyScheduling}. When a CES usage request is received, the CES manager generates a set of feasible schedules $S_n$ and then the best schedule, $s^*$, is chosen in line \ref{best}. \textcolor{black}{We note that our algorithm is equivalent to a posted price mechanism where all the options are enumerated with corresponding prices for each. Users simply examine their valuations for each feasible schedule, subtract the current cost of each schedule ($\tilde{p}_{ns}$), and choose the utility maximizing schedule (i.e., no complex optimization needed, they simply choose the highest value option). We note that the total price that the customer pays for their allocated schedule is calculated in line 9 of Algorithm 1 and is denoted as $\tilde{p}_{ns^*}$} After each request is scheduled or denied, the CES manager updates the dual variables with the new values for charging and discharging power as well as energy capacity (lines \ref{totals}-\ref{prices}).
\begin{algorithm}[]
\small
    \caption{\textsc{CommunityEnergyScheduling}}
    \label{algorithm}
    \begin{algorithmic}
    \STATE \textbf{Input:} $\hat{E}, \hat{P}_c, \hat{P}_d, L_{e,c,d}, U_{e,c,d}$
    \STATE \textbf{Output:} $x, p$
    \end{algorithmic}
    \begin{algorithmic}[1]
    \STATE Define the update functions $p(y(t))$ according to \eqref{eqn: energy function} - \eqref{eqn: discharging power function} for energy capacity, charging, and discharging.
    \STATE Initialize $x_{ns}=0$, $y_{e,c}(t)=0$, $u_n=0$.
    \STATE Initialize prices $p(0)$ according to \eqref{eqn: energy function} - \eqref{eqn: discharging power function}.
    \STATE \textbf{Repeat for all $N$ CES requests:}
    \STATE Request $n$ is received, generate feasible charging/discharging schedules $\mathcal{S}_n$
    \STATE Update dual variable $u_n$ according to \eqref{eqn: u_n}. \label{alg:utility}
    \IF{$u_n > 0$}
        \STATE \label{best} $(s^{\star}) =\argmax_{\mathcal{S}_n}\big\{v_{ns}$ $- \sum_{t\in[t_{ns}^-,t_{ns}^+]} \big( i_{nse}(t)p_{e}(t)$\\
        \vspace{3pt}
        \hspace{70pt} $ +i_{nsc}(t)p_{c}(t)
        -i_{nsc}(t)p_{d}(t)\big)\big\}$
        \vspace{3pt}
        \STATE $\widetilde{p}_{ns^{\star}} = \sum_{\mathcal{T}} \Big[ i_{nse}(t)p_e(t)
        + i_{nsc}(t)p_c(t)
        - i_{nsc}(t)p_d(t) \Big]$\\*
        \vspace{3pt}
        \STATE $x_{ns^{\star}}=1$ and $x_{ns}=0$ for all $s \neq s^{\star}$
        \STATE \label{totals} Update total demand $y(t)$ for energy capacity and charging power according to \eqref{eqn:y_e}-\eqref{eqn:y_c}. \label{alg:demand update}
        \STATE \label{prices} Update dual variables $p(y(t))$ for energy capacity, charging, and discharging according to \eqref{eqn: energy function} - \eqref{eqn: discharging power function}. \label{alg:price update}
    \ELSE
        \STATE $x_{ns}=0$, \hspace{3pt]}  $\forall$ $s\in\mathcal{S}_n$.
    \ENDIF 
    \IF{$\exists s^{\star}$ and $x_{ns^{\star}}=1$}
        \STATE Allocate request $n$ the energy capacity, charging power, and discharging power from schedule $s^*$.
        \STATE Request $n$ is fulfilled by schedule $s^*$ for the price of $\widetilde{p}_{ns^{\star}}$ to the requester.
    \ELSE
        \STATE Deny request $n$ from using the CES.
    \ENDIF
    \end{algorithmic}
\end{algorithm}

The three main benefits are as follows: 1) the online scheduling heuristic ensures that the utility gained from each scheduled request is positive for the user, 2) the online scheduling heuristic filters out low value charging and discharging requests in order to prevent the CES from being overused, and 3) the online scheduling heuristic promotes diverse charging and discharging schedules to take advantage of charge/discharge cancellations as mentioned in Section \ref{subsection: charging and discharging cancellation}. The underlying framework of the dual variable update heuristic is similar to that of \cite{IaaS}, in which the authors present an auction mechanism for optimizing the usage of computer hardware at data centers for cloud computing.

In our online CES scheduling heuristic, we expand upon the specialized functions proposed in \cite{IaaS} that approximate the optimal dual variables in an online fashion. These dual variable functions depend on the amount of energy capacity, charging power, and discharging power that is reserved at a future time $t$. The update functions increase slowly at first then increase rapidly as the CES power and capacity limits are approached. Additionally, when the power and capacity limits are met, the dual variable update functions ensure that no more schedules will be allocated by outputting dual variables high enough to ensure no schedule yields positive utility, thus enforcing the hard capacity and power limits. The specialized function to update the dual variable associated with the energy capacity of the CES is as follows:
\begin{align}
\label{eqn: energy function}
    p_e(t) = \Big(\frac{L_e}{6}\Big) \Big(\frac{6U_e}{L_e}\Big)^{\frac{y_e(t)}{\hat{E}}}, \;y_e(t)\in [0, \hat{E}], 
\end{align}
where $U_e$ and $L_e$ correspond to the maximum and minimum value per kWh of energy capacity per time unit, respectively, across all requests. We note that the CES manager does require knowledge of $U_e$ and $L_e$ beforehand to calculate initial values for the dual variables and to ensure limits are not breached. The maximum and minimum valuations are calculated as follows:
\begin{align}
    L_e = \min_{n\in\mathcal{N},s\in\mathcal{S}_n} \frac{v_{ns}}{3\sum_{t\in [t_s^-, t_s^+]} i_{nse}(t)},
\end{align}

\begin{align}
    U_e = \max_{n\in\mathcal{N},s\in\mathcal{S}_n,t\in\mathcal{T}} \frac{v_{ns}}{i_{nse}(t)}, \; i_{nse}(t) > 0.
\end{align}
In addition to the energy capacity's dual variable update function in \eqref{eqn: energy function}, the dual variables for the charging and discharging power of the CES also require update functions:
\begin{align}
\label{eqn: charging power function}
    p_c(t) = \Big(\frac{L_c}{6}\Big) \Big(\frac{6U_c}{L_c}\Big)^{\frac{y_c(t)}{\hat{P}_{c}}},\;y_{c}(t) \in [-\hat{P}_{d}, \hat{P}_{c}],
\end{align}

\begin{align}
\label{eqn: discharging power function}
    p_d(t) = \Big(\frac{L_d}{6}\Big) \Big(\frac{6U_d}{L_d}\Big)^{\frac{-y_c(t)}{\hat{P}_{d}}}, \;y_{c}(t) \in [-\hat{P}_{d}, \hat{P}_{c}].
\end{align}
We note that the dual variable update functions for charging and discharging power, \eqref{eqn: charging power function} and \eqref{eqn: discharging power function}, are similar to the energy capacity dual variable function \eqref{eqn: energy function} except for the domain. The energy capacity function's input values, $y_e(t)$, are nonnegative and less than $\hat{E}$. The charging and discharging functions' input values can be negative and are within the range $y_{c}(t) \in [-\hat{P}_{d}, \hat{P}_{c}]$. 
With the 3 dual variable update functions \eqref{eqn: energy function}, \eqref{eqn: charging power function}, and \eqref{eqn: discharging power function}, we now have the means to calculate estimates for the optimal dual variables in order to solve \eqref{eqn: u_n} in an online fashion (i.e., at the reception of each request to use the CES). The full procedure can be seen in Algorithm 1 \textsc{\textsc{CommunityEnergyScheduling}}.


The heuristic presented in \textsc{CommunityEnergyScheduling} attempts to solve an online scheduling problem without full knowledge of the sequence of requests. As stated before, we are able to compare the total welfare generated from our online heuristic to the total welfare generated by an omniscient offline CES manager. The comparison that we make is in the form of a competitive ratio. An online heuristic is said to be $\alpha$-competitive when the ratio of welfare generated by the omniscient offline solution to the welfare generated by the online heuristic is bounded by $\alpha\geq1$. \textcolor{black}{The competitive ratio, $\alpha$, is defined as $OPT/ALG_{worstcase} \geq 1$, where $OPT$ is the welfare generated by the offline optimal solution and $ALG_{worstcase}$ is the worst-case welfare generated by the online algorithm. A value of 1 means the algorithm performs optimally and higher values of $\alpha$ indicate worse performance.} In this work, we build upon results from \cite{IaaS} (and previous work \cite{8814409, ntucker_ITSC, 8721152}) and present a competitive ratio that accounts for the cancellation of  complimentary resources (e.g., charging and discharging power, which previous works could not account for). For the following results, we assume that each CES request utilizes a small amount of the charging/discharging power and energy capacity of the CES to ensure that one schedule cannot prohibit numerous future schedules and that the ratios of users' maximum valuation to minimum valuation for charging and discharging power are equal, i.e., $\frac{U_c}{L_c}=\frac{U_d}{L_d}=\frac{U_{c,d}}{L_{c,d}}$ (to yield a singular $\alpha_{c,d}$ for both the charging and discharging of the CES).
\begin{theorem}
\label{theorem: competitive ratio}
    The community energy storage system's schedules generated by \textsc{CommunityEnergyScheduling} in Algorithm 1 are $\alpha$-competitive in welfare over N usage requests where $\alpha = \max\{\alpha_e, \alpha_{c,d}\}$ and $\alpha_e$ and $\alpha_{c,d}$ are defined as follows:
    \begin{align}
        &\nonumber\alpha_e= 2\ln{\Big(\frac{6 U_e}{L_e}\Big)},\\
        &\nonumber\alpha_{c,d}= 2\ln{\Big(\frac{6 U_{c,d}}{L_{c,d}}\Big)}.
    \end{align}
\end{theorem}

\noindent\textit{Proof.} The full proof can be found in the Appendix. The CES system has two limited resources that must be scheduled, the energy capacity and the charging power. From Lemmas 1-5 (in the Appendix), there are independent welfare guarantees $\alpha_e$ and $\alpha_{c,d}$ for the energy capacity as well as the charging/discharging power schedules of the CES, respectively. To find the all-encompassing $\alpha$ for the entire CES, we take the maximum between $\alpha_{e}$ and $\alpha_{c,d}$ to yield the bound that accounts for both the energy capacity and the charging and discharging power. \hfill $\square$

\section{Numerical Results}
\label{sec: numerical results}

In the following, we present two different numerical results to showcase our heuristic. First, we describe an example system in Section \ref{subsec: intuitive example} which explicitly details the CES requests' valuations and compares the social welfare generated from our proposed heuristic to the optimal offline case as the users submit requests to use the shared battery. We then present a larger case study for a shared battery system serving commercial customers in California in Section \ref{subsec: LA case study}.

\subsection{Intuitive Example}
\label{subsec: intuitive example}

\begin{figure*}[]
    \centering
    \includegraphics[width=\textwidth]{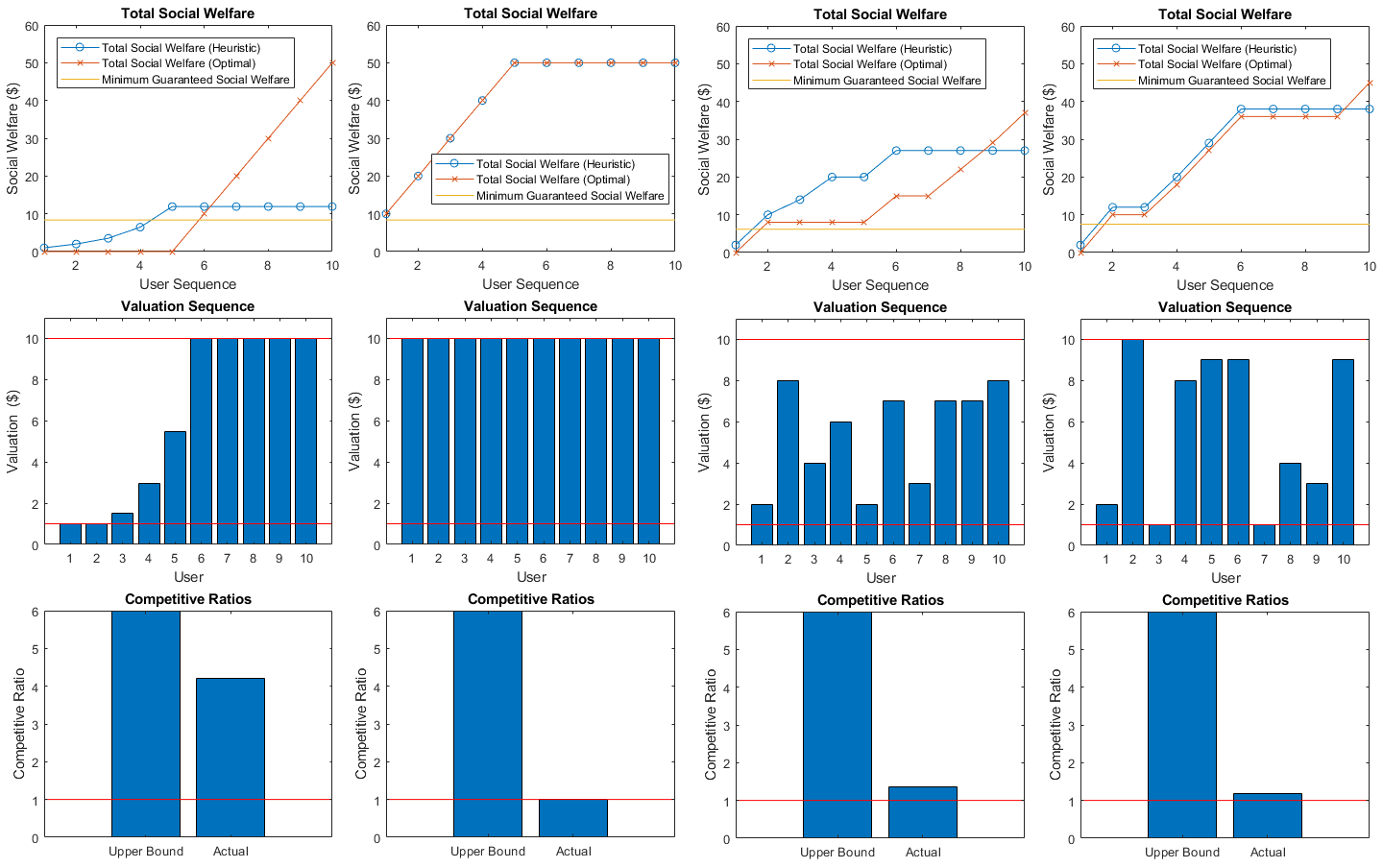}
    \caption{Simple example system with 10 user requests to use a 5kWh shared battery system. Each column corresponds to a different arrival sequence of user valuations. }
    \label{fig: intuitive}
\end{figure*}

In order to showcase the details of our heuristic, we make use of a specific example setup. Namely, we consider a shared battery system that has a maximum charging rate of 5kW, maximum storage of 5kWh, and a maximum discharging rate of 5kW. Furthermore, we consider 10 unique users who want to purchase the exact same charging, storage, and discharging schedule and are interested in no other schedules (the specific schedule of interest is to charge 1kW from 8-9am, store 1kWh from 9-10am, and discharge 1kW from 10-11am). These 10 unique users arrive sequentially one after another and submit their bids to purchase the charging, storage, and discharging schedule (the specific times that each user submits their request are irrelevant as long as they are submitted sequentially and all before 8am). For this example, we assume the users' valuations are within $\$1$ and $\$10$. Due to the constraints of the shared battery system, it is clear that only 5 of the 10 users will be able to use the  battery for that specific charging, storage, and discharging schedule. In the offline case, the optimal solution yielding maximal social welfare will select the 5 users with the highest valuations to use the shared battery. However, since the users submit their bids sequentially and their valuations are unknown a priori, our heuristic attempts to emulate the offline solution via dynamic prices that increase as the battery usage increases, thus filtering out users with low valuations.

In Figure \ref{fig: intuitive}, we present the results of 4 different user valuation sequences (each column corresponds to a different sequence of user valuations). Row 1 presents the social welfare results of our heuristic and the optimal offline solution. Row 2 presents the users' valuations (in order). Row 3 presents the competitive ratio upper bound from our theoretical results in addition to the actual competitive ratio for that column's request sequence.  From left to right: Column 1 portrays the worst case user valuation sequence. This is because each user's valuation was carefully selected to equal the current price of the schedule generated by our heuristic (i.e., each of the first 5 users have the minimum valuations that our pricing heuristic will accept while the last 5 users have the maximum valuation, thus leading to the worst possible competitive ratio). Note that the actual competitive ratio in this case is still below the theoretical upper bound. Column 2 portrays one of the many valuation sequences where the heuristic matches the offline optimal solution (i.e., competitive ratio = 1). Columns 3 and 4 present randomly generated valuation sequences (i.e., user valuations were drawn from a uniform distribution between \$1 and \$10) to showcase that our heuristic often yields competitive ratios close to 1.

\textcolor{black}{Additionally, we compare each of the 4 arrival sequences in Fig. \ref{fig: intuitive} to a First-Come-First-Serve (FCFS) heuristic that is the status quo scheduling method for any new CES implementation. Table \ref{table: FCFS} presents the percentage of the offline optimal welfare that is generated by both our Algorithm 1 and a FCFS heuristic.}
\begin{table}[h!]
\color{black}
\small
\begin{center}
\begin{tabular}{||c c c c c||} 
 \hline
 & Sequence 1 & Sequence 2& Sequence 3 & Sequence 4 \\ [0.5ex] 
 \hline\hline
 ALG1 & 24\% & 100\% & 73\% & 87\%\\ 
 \hline
 FCFS & 24\% & 100\% & 57\% & 66\%\\
 \hline
\end{tabular}
\end{center}
\caption{\small\textcolor{black}{Percentage of offline optimal welfare generated by Algorithm 1 and First-Come-First-Serve.}}
\label{table: FCFS}
\end{table}

\begin{figure*}[ht!]
    \centering
    \includegraphics[width=\textwidth]{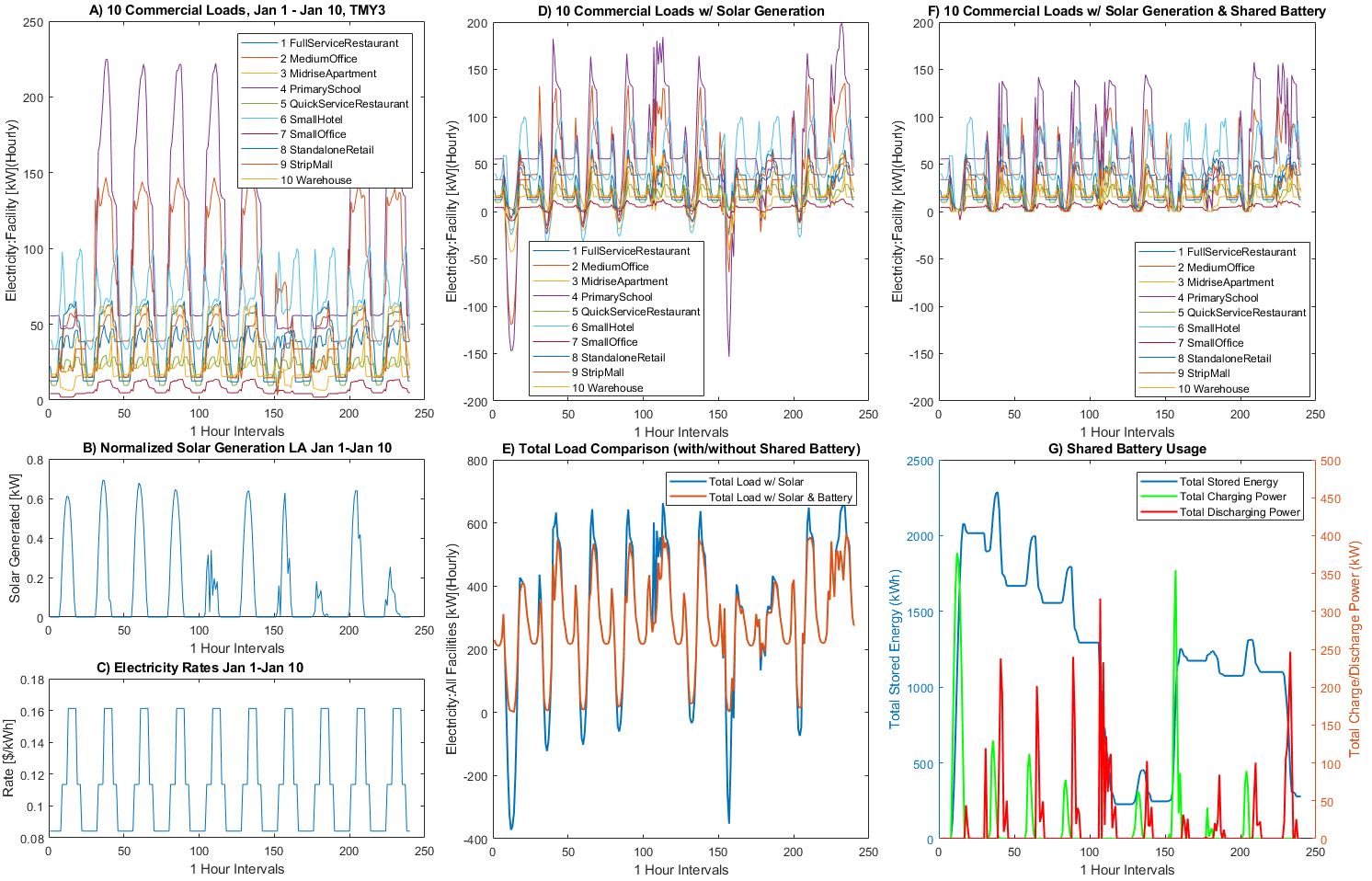}
    \caption{Simulation Results for California Test Case.}
    \label{fig: LA results}
\end{figure*}

\subsection{California Case Study}
\label{subsec: LA case study}

In this section we present results from a community energy storage system in California. Specifically, there are 10 loads (presented in Fig. \ref{fig: LA results}.A) sourced from a commercial building load dataset \cite{Commercial}. The publicly accessible dataset \cite{Commercial} contains hourly load profile data for commercial building types and residential buildings in all TMY3 locations in the United States. The Typical Meteorological Year 3 (TMY3) provides one year of hourly data that best represents median weather conditions over a multiyear period for a particular location. Across the 10 day time span, January 1st - January 10th, we also assume that each load is equipped with behind-the-meter solar generation that they would like to charge and discharge the CES with. The normalized solar generation for the California location \cite{Solar_PVWATTS} is presented in Fig. \ref{fig: LA results}.B. We assume that each building is equipped with solar generation capacity to fulfill 80$\%$ of their peak load at maximum rating. Fig. \ref{fig: LA results}.D presents the 10 loads once the solar generation is subtracted. Note that negative power means that the location is producing more power than is being consumed. We assume that all 10 buildings are able to use a 2500 kWh community energy storage system with maximum charge and discharge rates of 500 kW. Furthermore, we assume that the 10 buildings are connected to the local grid and pay time-of-use electricity rates \cite{electricity_rates} for energy that is not provided by their solar generation. The electricity rate used is the PG\&E E-19 structure for buildings $<$1000 kW max demand and is shown in Fig. \ref{fig: LA results}.C. For the purposes of this work, we do not consider net energy metering for the locations injecting excess solar generation back into the local grid as sending excess energy to the CES is preferred. 

As noted in Section \ref{subsection: energy storage schedules}, the incentive for the buildings to use the CES comes from storing excess solar generation and using it at a later time. As such, whenever a building detects that it is producing more power than it needs, it submits a request to store that excess power in the CES. Specifically, on an hour-by-hour basis, each location submits requests to store their excess energy in the CES. In order to accomodate this, the CES manager limits the number of feasible charging/storage/discharging schedules to 96 for each request. Namely, all the excess generation that the building wants to inject into the CES during hour $t$ must be discharged at the same rate during a future hour in the range $[t, t+96]$ (i.e., in the next 4 days). The valuation for each of the 96 schedules is calculated via equation \eqref{eqn: valuation} (i.e., the predicted cost savings from using stored energy versus purchasing energy from the grid).

As portrayed in Fig. \ref{fig: LA results}.E, the total load of all 10 buildings is greatly affected by the CES usage. Specifically, the cumulative load no longer goes negative (the red curve in 2.E), meaning that the buildings are not injecting solar back into the local grid. Instead, they are storing that power and using it to reduce peak demands at later times. This helps reduce electricity costs for the buildings in addition to reducing the stress on the load grid from injecting the excess solar generation. Last, in Fig. \ref{fig: LA results}.G, we present the charging, discharging, and total energy stored in the CES throughout the 10 days. 

\subsection{Additional Case Study}
\label{subsec: additional case study}
\begin{figure*}[]
    \centering
    \includegraphics[width=0.9\textwidth]{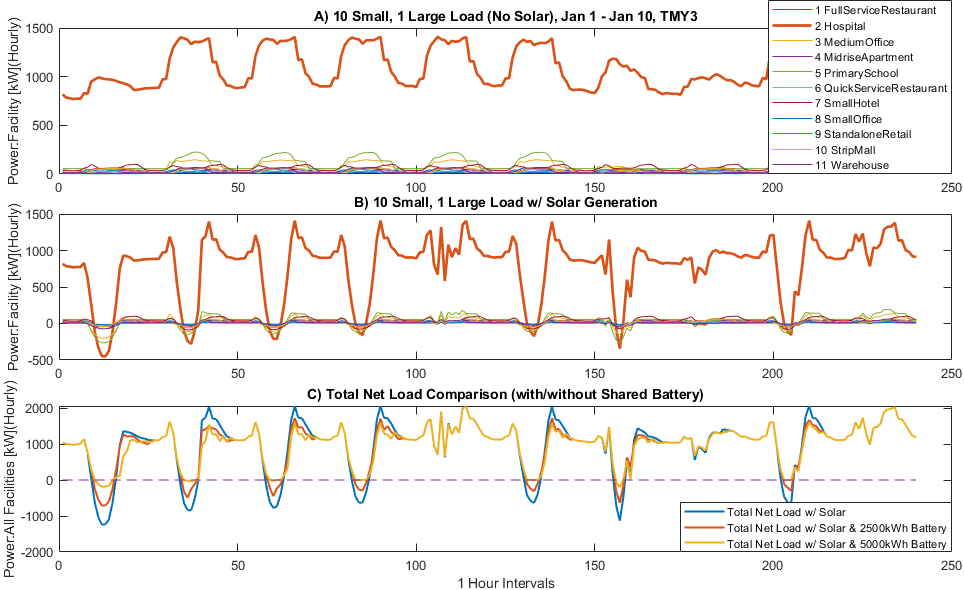}
    \caption{\textcolor{black}{Load profiles for energy community and a large hospital.} }
    \label{fig:hospital}
\end{figure*}

\textcolor{black}{In Figure \ref{fig:hospital}, we present results for the same energy community as in Section \ref{subsec: LA case study}; however, we include a large hospital as one of the loads (hospital also from dataset \cite{Commercial}). As seen in plots A) and B) of Figure \ref{fig:hospital}, the load (A) and net load (B) of the hospital are significantly larger than the other 10 loads. While it is possible that the large hospital might dominate usage of the CES, due to the fact that valuations are bounded per time slot for all users and users only purchase CES schedules if they are cheaper than the current grid electricity prices, all users end up with a fair chance at CES usage. Additionally, the smaller users have slightly different load patterns than the large hospital yielding many charge/discharge cancellations. Moreover, if the smaller users submit their CES requests before the hospital, they could exclude the hospital's large charge/discharge requests due to the capacity and power constraints. Last, we note that our results can begin to degrade if the size of the CES is not large enough to adequately supply all of the users. Specifically, in the case of the 10 small users and the large hospital, we assume that any CES manager would install a large enough battery to give all users a fair chance at usage. In plot C) of Figure \ref{fig:hospital}, we show the total net load of all 11 users with no CES, with a 2500kWh battery (same as in Section \ref{subsec: LA case study})  and a 5000kWh battery. We note that the 2500kWh battery was not large enough to store all of the excess solar generation; however, the 5000kWh battery nearly stored all excess solar generation.}

\section{Conclusion}
We presented a solution to the problem facing the manager of a community energy storage system attempting to schedule the charging/discharging/energy storage of the system. We presented an online heuristic that updates dual variables in real-time as a solution to the problem. The heuristic acts as a pricing mechanism to ensure the CES yields positive utility and promotes charge and discharge cancellations to reduce the CES's usage at popular times. The heuristic is able to handle the inherently stochastic nature of the requests to charge and discharge from the CES (stemming from weather uncertanties and randomness in users' electricity usage patterns). The heuristic can handle adversarially chosen request sequences and will always yield total welfare within a factor of $\frac{1}{\alpha}$ of the offline optimal welfare. An intuitive example was presented to showcase the heuristics performance for various request sequences and a larger case study was presented for 10 commercial buildings sharing a CES.

\bibliographystyle{IEEEtran}
\bibliography{references}



\textcolor{black}{
\section*{Appendix}
\begin{flalign}
    &\mathcal{N} &&\text{Set of CES requests indexed by }n=1,\dots,N \nonumber \\[-2.0pt]
    &\mathcal{T} &&\text{Set of time intervals indexed by }t=1,\dots,T \nonumber \\[-2.0pt]
    &\mathcal{S}_n && \text{Set of feasible schedules that satisfy request }n \nonumber \\[-2.0pt]
    &\hat{E}  &&\text{CES's max energy capacity } \nonumber \\[-2.0pt]
    &\hat{P}_c  &&\text{CES's max charging power} \nonumber \\[-2.0pt]
    &\hat{P}_d  &&\text{CES's max discharging power } \nonumber \\[-2.0pt]
    &t_n^-  &&\text{Request }n\text{'s schedule start time} \nonumber \\[-2.0pt]
    &t_{ns}^+  &&\text{Request }n\text{'s schedule option }s\text{'s end time} \nonumber \\[-2.0pt]
    &i_{nsc}(t) && \text{CES charging power schedule for request $n$ in $s$}\nonumber \\*[-2.0pt]
    &i_{nse}(t) && \text{CES energy capacity schedule for request $n$ in $s$}\nonumber \\*[-2.0pt]
    &v_{ns}  &&\text{Request }n\text{'s valuation for option schedule }s \nonumber \\[-2.0pt]
    &p_{grid}(t)  &&\text{Electricity rate from local grid at time }t \nonumber \\[-2.0pt]
    &x_{ns} && \text{Binary assignment variable for request $n$ for}\nonumber \\*[-2.0pt]
    & && \text{schedule option $s$} \nonumber \\[-2.0pt]
    &\tilde{p}_{ns} && \text{Payment for request $n$ for schedule option $s$}\nonumber \\*[-2.0pt]
    &y_{e}(t) && \text{Total CES energy capacity  reserved  at time $t$}\nonumber \\*[-2.0pt]
    &y_{c}(t) && \text{Total CES charging/discharging  reserved  at time $t$}\nonumber \\*[-2.0pt]
    &u_n &&\text{Utility for requester $n$ from the CES system} \nonumber \\[-2.0pt]
    &p_e(t) &&\text{CES energy capacity resource price at time $t$} \nonumber \\[-2.0pt]
    &p_c(t) &&\text{CES charging power resource price at time $t$} \nonumber \\[-2.0pt]
    &p_d(t) &&\text{CES discharging power resource price at time $t$} \nonumber \\[-2.0pt]
    &f^*(\cdot)  &&\text{Fenchel conjugate of a cost function/constraint} \nonumber \\[-2.0pt]
    &L_{e,c,d}  &&\text{Lower bound on valuations per resource} \nonumber \\[-2.0pt]
    &U_{e,c,d}  &&\text{Upper bound on valuations per resource} \nonumber
\end{flalign}
}

\setcounter{theorem}{0}
\begin{theorem}
\label{theorem: competitive ratio}[Repeated.]
    The community energy storage system's schedules generated by \textsc{CommunityEnergyScheduling} in Algorithm 1 are $\alpha$-competitive in welfare over N usage requests where $\alpha = \max\{\alpha_e, \alpha_{c,d}\}$ and $\alpha_e$ and $\alpha_{c,d}$ are defined as follows:
    \begin{align}
        &\nonumber\alpha_e= 2\ln{\Big(\frac{6 U_e}{L_e}\Big)},\\
        &\nonumber\alpha_{c,d}= 2\ln{\Big(\frac{6 U_{c,d}}{L_{c,d}}\Big)}.
    \end{align}
\end{theorem}

\noindent The proof of Theorem \ref{theorem: competitive ratio} requires the following Definition and Lemmas.

\noindent\textbf{Definition 1. } (From \cite{IaaS}) \textit{The Differential Allocation-Payment Relationship for a given parameter $\alpha \geq 1$ is:}
\begin{align}
\label{eqn:diffalloc}
    \big(p(t) - f'(y(t))\big) \text{d}y(t) \geq \frac{1}{\alpha(t)} f^{*'}(p(t)) \text{d}p(t)
\end{align}
for all $t\in[0,T]$ and for all shared resources (energy capacity, discharging power, and charging power) where $f'(y(t))$ is the derivative of an operational cost function and $f^{*'}(p(t))$ is the corresponding Fenchel conjugate's derivative.

\vspace{2pt}    
In the following, let $i_{nsd}(t)$ be the CES \textit{discharging power profile} for request $n$ in feasible schedule $s$ at time $t$. We note that the charging power profile $i_{nsc}(t)|_{t=1,\dots,T}$ and the discharging power profile $i_{nsd}(t)|_{t=1,\dots,T}$ are negatives of one another. We add this variable to separate the charging power dual variable updates from the discharging power dual variable updates for ease of exposition. Specifically, let $i_{nsd}(t) = -i_{nsc}(t), \forall t $. Additonally, we add the variable $y_d(t)$ to denote the total discharging power at time $t$. Specifically, let $y_d(t) = -y_c(t), \forall t$. The variable $y_d(t)$ can be calculated similarly to $y_c(t)$ in \eqref{eqn:y_c} as: $y_{d}(t)=\sum_{\mathcal{N},\mathcal{S}_n} i_{nsd}(t) x_{ns}$.
    
\begin{lemma}
\label{Background 2}
(From \cite{IaaS}) If the \textit{Differential Allocation-Payment Relationship} holds for $\alpha\geq1$, then each energy storage request $n$ and the chosen charge/discharge schedule $s_n^{\star}$ satisfy the following:
\end{lemma}
\begin{align*}
    \widetilde{p}_{ns^{\star}} - \sum_{{t\in[t_{ns}^-,t_{ns^{\star}}^+]}} &\bigg( \Delta f_e(y_e(t))^{(n,n-1)}
    + \Delta f_c(y_c(t))^{(n,n-1)}\\
    &+ \Delta f_d(y_d(t))^{(n,n-1)}
    \bigg)\\
    &\geq \frac{1}{\alpha}(D^n - D^{n-1} - u_n)
\end{align*}

\noindent where 

\begin{align*}
    \Delta f_e(y_e(t))^{(n,n-1)} = f_e(y_e(t))^{(n)} - f_e(y_e(t))^{(n-1)}
\end{align*}

\begin{align*}
    \Delta f_c(y_c(t))^{(n,n-1)} = f_c(y_c(t))^{(n)} - f_c(y_c(t))^{(n-1)}
\end{align*}

\begin{align*}
    \Delta f_d(y_d(t))^{(n,n-1)} = f_d(y_d(t))^{(n)} - f_d(y_d(t))^{(n-1)}
\end{align*}

\begin{align*}
    \widetilde{p}_{ns^{\star}} = \sum_{\mathcal{T}} \Big[ i_{nse}(t)p_e(t)
    + i_{nsc}(t)p_c(t)
    + i_{nsd}(t)p_d(t) \Big].
\end{align*}

\noindent\textit{Proof of Lemma \ref{Background 2}}: We expand out $D^n-D^{n-1} =$
\begin{align*}
    u_n + \sum_{t\in[t_n^-,t_{ns^{\star}}^+]}  &\bigg( \Delta f_e^*(p_e(t))^{(n,n-1)}
    + \Delta f_c^*(p_c(t))^{(n,n-1)}\\
    &+ \Delta f_d^*(p_e(t))^{(n,n-1)}\bigg)
\end{align*}
\noindent where 
\begin{align*}
    \Delta f_e^*(y_e(t))^{(n,n-1)} = f_e^*(y_e(t))^{(n)} - f_e^*(y_e(t))^{(n-1)}
\end{align*}

\begin{align*}
    \Delta f_c^*(y_c(t))^{(n,n-1)} = f_c^*(y_c(t))^{(n)} - f_c^*(y_c(t))^{(n-1)}
\end{align*}

\begin{align*}
    \Delta f_d^*(y_d(t))^{(n,n-1)} = f_d^*(y_d(t))^{(n)} - f_d^*(y_d(t))^{(n-1)}.
\end{align*}
The lemma follows by summing the \textit{Differential Payment-Allocation Relationship} over all shared resources (energy capacity, discharge power, and charge power) and over the entire time period.\hfill $\square$

\begin{lemma}
\label{Background 3}
(From \cite{IaaS}) If the Differential Allocation-Payment Relationship holds for $\alpha\geq1$ then $P^n-P^{n-1} \geq \frac{1}{\alpha} (D^n - D^{n-1}) $ for all $n$.
\end{lemma}
\vspace{2ex}

\noindent \textit{Proof of Lemma \ref{Background 3}}: If energy storage request $n$ is denied for all schedules $s\in\mathcal{S}_n$, then $P^n-P^{n-1} = D^n - D^{n-1}=0$. Otherwise, the change of the primal objective is:
\begin{align*}
    P^n - P^{n-1} = &v_{ns^{\star}} - \sum_{t\in[t_n^-,t_{ns^{\star}}^+]} \Big( \Delta f_e(y_e(t))^{(n,n-1)}\\
    &+ \Delta f_c(y_c(t))^{(n,n-1)} + \Delta f_d(y_d(t))^{(n,n-1)}\Big)
\end{align*}
where $v_{ns^{\star}} = u_n + \widetilde{p}_{ns^{\star}}$. By Lemma \ref{Background 2}, we get that
\begin{align*}
    P^n - P^{n-1} \geq u_n + \frac{1}{\alpha}(D^n - D^{n-1} -u_n).
\end{align*}
With $u_n\geq 0$ and $\alpha\geq1$, then $P^n-P^{n} \geq \frac{1}{\alpha} (D^n - D^{n-1})\; \forall n\in\mathcal{N} $.\hfill$\square$
    
    \begin{lemma}
\label{Background 1}
(From \cite{IaaS}) If there is a constant $\alpha \geq 1$ such that the incremental increase of the primal and dual objective values differ by at most an $\alpha$ factor, i.e., $P^n - P^{n-1} \geq \frac{1}{\alpha}(D^n - D^{n-1})$, for every energy storage request $n$, then the heuristic is $2\alpha$-competitive.
\end{lemma}
\noindent\textit{Proof of Lemma \ref{Background 1}}: Summing up the inequality at each step $n$, we have
\begin{align*}
    P^N &= \sum_n(P^n - P^{n-1})\\
    &\geq \frac{1}{\alpha} \sum_n (D^n - D^{n-1})\\
    &=\frac{1}{\alpha}(D^N-D^0).
\end{align*}
Now, we use the fact that the initial primal value is $P^0=0$ and by weak duality, $D^N \geq OPT$. Next, we assume $D^0\leq\frac{1}{2}OPT$, we have that $P^N \geq \frac{1}{2\alpha}OPT$. Thus, the online heuristic is $2\alpha$-competitive. $\hfill\square$ 


\begin{lemma}
\label{lemma energy capacity}
The online pricing heuristic \eqref{eqn: energy function} is $\alpha_e$-competitive in welfare generated from the scheduling of energy capacity in the shared battery where
\begin{align}
\nonumber
    \alpha_e= 2\ln{\Big(\frac{6 U_e}{L_e}\Big)}.
\end{align}
\end{lemma}
\vspace{1ex}
\noindent \textit{Proof of Lemma \ref{lemma energy capacity}}: We will show that the pricing heuristic in \eqref{eqn: energy function} satisfies the \textit{Differential Payment-Allocation Relationship} in equation \eqref{eqn:diffalloc} with parameter $\alpha_e$. Then the rest of the Lemma follows from Lemmas \ref{Background 2}, \ref{Background 3}, and \ref{Background 1}.

The scheduled energy capacity of the shared battery has no cost to the battery manager but cannot exceed the total capacity limit of the battery $\hat{E}$ (in other terms, the cost function $f_e(y_e(t))$ for the energy capacity can be seen as a zero-infinite step function with the step occurring right after $\hat{E}$). Furthermore, the pricing function \eqref{eqn: energy function} never allows $y_e(t)$ to exceed $\hat{E}$ so the derivative $f_e'(y_e(t))=0$ while $y_e(t)\leq \hat{E}$ (and $y_e(t)\leq \hat{E} \;\;\forall t$ due to \eqref{eqn: energy function} outputting prices too high for any user once the used battery capacity is at $\hat{E}$). Next, the derivative of the Fenchel conjugate \eqref{eqn: fenchel energy} for the energy capacity is as follows:$f_e^{*'}(p_e(t)) = \hat{E}$.

The derivative of the proposed pricing function \eqref{eqn: energy function} is

\begin{align}
    \nonumber\text{d}p_e(t) = &\Big(\frac{L_e}{6 \hat{E}}\Big)\Big(\frac{6 U_e}{L_e}\Big)^{\frac{y_e(t)}{\hat{E}}}
    \ln{\Big(\frac{6 U_e}{L_e}\Big)}\text{d}y_e(t).
\end{align}

After inserting $f_e'(y_e(t))$, $ f_e^{*'}(p_e(t))$, and $\text{d}p_e(t)$ in \eqref{eqn:diffalloc}, we can show that the Differential Allocation-Payment Relationship holds when choosing $\alpha = \alpha_e= \ln{\Big(\frac{6 U_e}{L_e}\Big)}$. Because \eqref{eqn:diffalloc} holds for the dual variable update function, cost function, and Fenchel conjugate, the remainder of the proof follows from Lemmas \ref{Background 2}, \ref{Background 3}, and \ref{Background 1}.\hfill$\square$

\noindent\textbf{Definition 2. }  The \textit{Generalized Differential Allocation-Payment Relationship} for the payment and remuneration of two coupled resources (resources $a$ and $b$) for a given parameter $\alpha \geq 1$ is:
\begin{align}
\label{eqn:gendiffalloc}
    &\nonumber\big[p_a(t) - f_a'(y_a(t))\big] \text{d}y_a(t) 
    +\big[p_b(t) - f_b'(y_b(t))\big] \text{d}y_b(t)\\
    &\geq \frac{1}{\alpha(t)} \Big[ f_a^{*'}(p_a(t)) \text{d}p_a(t) + f_b^{*'}(p_b(t)) \text{d}p_b(t)\Big]
\end{align}
for all $t\in[0,T]$ where $f'(y(t))$ is the derivative of an operational cost function and $f^{*'}(p(t))$ is the corresponding Fenchel conjugate's derivative.

\begin{lemma}
\label{lemma charge/discharge power}
The online pricing heuristics \eqref{eqn: charging power function} and \eqref{eqn: discharging power function} are $\alpha_{c,d}$-competitive in welfare generated from the scheduling of charging and discharging power in the shared battery where
\begin{align}
\nonumber
    \alpha_{c,d}= 2\ln{\Big(\frac{6 U_{c,d}}{L_{c,d}}\Big)}.
\end{align}
\end{lemma}
\vspace{1ex}
\noindent \textit{Proof of Lemma \ref{lemma charge/discharge power}}:
We will show that the pricing heuristics in \eqref{eqn: charging power function} and \eqref{eqn: discharging power function} satisfy a \textit{Generalized Differential Payment-Allocation Relationship} that handles both payments and remunerations of coupled resources  such as charging and discharging power with parameter $\alpha_{c,d}$. Then the rest of the Lemma follows from Lemmas \ref{Background 2}, \ref{Background 3}, and \ref{Background 1}.  

The proof follows similarly to that of Lemma \ref{lemma energy capacity}. Both the charging power and discharging power resources have zero-infinite step functions for their operational cost functions with the step occurring at the max charging power  $\hat{P}_c$ and discharging power $\hat{P}_d$, respectively.  Furthermore, each of the pricing functions \eqref{eqn: charging power function} and \eqref{eqn: discharging power function} never allow $y_c(t)$ and $y_d(t)$ to exceed $\hat{P}_c$ and $\hat{P}_d$, respectively. Thus, $f_c'(y_c(t))=0$ and $f_c'(y_d(t))=0$. Next, the derivatives of the Fenchel conjugates are $f_c^{*'}(p_c(t)) = \hat{P}_{c}$ and $f_d^{*'}(p_d(t)) = \hat{P}_{d}$. The derivatives of the charging power pricing function \eqref{eqn: charging power function} and the discharging power pricing function \eqref{eqn: discharging power function} are as follows:
\begin{align*}
    &\nonumber\text{d}p_c(t) = \Big(\frac{L_c}{6 \hat{P}_c}\Big)\Big(\frac{6 U_c}{L_c}\Big)^{\frac{y_c(t)}{\hat{P}_c}}
    \ln{\Big(\frac{6 U_c}{L_c}\Big)}\text{d}y_c(t),\\
    &\nonumber\text{d}p_d(t) = \Big(\frac{L_d}{6 \hat{P}_d}\Big)\Big(\frac{6 U_d}{L_d}\Big)^{\frac{y_d(t)}{\hat{P}_d}}
    \ln{\Big(\frac{6 U_d}{L_d}\Big)}\text{d}y_d(t).
\end{align*}
After inserting $f_c'(y_c(t))$, $ f_c^{*'}(p_c(t))$, $f_d'(y_d(t))$, $ f_d^{*'}(p_d(t))$, $\text{d}p_c(t)$, and $\text{d}p_d(t)$ in \eqref{eqn:gendiffalloc}, the relationship is as follows:
\begin{align*}
    &\Big(\frac{L_c}{6}\Big) \Big(\frac{6U_c}{L_c}\Big)^{\frac{y_c(t)}
    {\hat{P}_{c}}}\text{d}y_c(t)
    + \Big(\frac{L_d}{6}\Big) \Big(\frac{6U_d}{L_d}\Big)^{\frac{y_d(t)}{\hat{P}_{d}}}
    \text{d}y_d(t)\\*
    &\geq 
    \frac{1}{\alpha(t)}\Bigg[
    \Big(\frac{L_c}{6}\Big)\Big(\frac{6 U_c}{L_c}\Big)^{\frac{y_c(t)}{\hat{P}_c}}
    \ln{\Big(\frac{6 U_c}{L_c}\Big)}\text{d}y_c(t)\\*
    &\hspace{40pt}+ \Big(\frac{L_d}{6}\Big)\Big(\frac{6 U_d}{L_d}\Big)^{\frac{y_d(t)}{\hat{P}_d}}
    \ln{\Big(\frac{6 U_d}{L_d}\Big)}\text{d}y_d(t)\Bigg].
\end{align*}
Now, using the assumption that the ratios of users' maximum valuation to minimum valuation for charging and discharging are equal, i.e., $\frac{U_c}{L_c}=\frac{U_d}{L_d}=\frac{U_{c,d}}{L_{c,d}}$, the relationship can be simplified to:
\begin{align*}
    &\Bigg[\Big(\frac{L_c}{6}\Big) \Big(\frac{6U_{c,d}}{L_{c,d}}\Big)^{\frac{y_c(t)}
    {\hat{P}_{c}}}\text{d}y_c(t)
    + \Big(\frac{L_d}{6}\Big) \Big(\frac{6U_{c,d}}{L_{c,d}}\Big)^{\frac{y_d(t)}{\hat{P}_{d}}}
    \text{d}y_d(t)\Bigg]\\*
    &\geq 
    \frac{\ln{\Big(\frac{6 U_{c,d}}{L_{c,d}}\Big)}}{\alpha(t)}
    \times\\*
    &\Bigg[
    \Big(\frac{L_c}{6}\Big)\Big(\frac{6 U_{c,d}}{L_{c,d}}\Big)^{\frac{y_c(t)}{\hat{P}_c}}
    \text{d}y_c(t)
    + \Big(\frac{L_{c}}{6}\Big)\Big(\frac{6 U_{c,d}}{L_{c,d}}\Big)^{\frac{y_d(t)}{\hat{P}_d}}
    \text{d}y_d(t)\Bigg].
\end{align*}
The bracketed term that is shared on the LHS and the RHS represents the total payment and remuneration for charging/discharging at a given time $t$. To simplify this relationship further, there are 3 cases: 1) when the payment is greater than the remuneration and the bracketed term is positive, 2) when the payment is less than the remuneration and the bracketed term is negative, and 3) when the payment is equal to the remuneration and the bracketed term is zero.

In case 1, the relationship simplifies to $\alpha\geq\ln\Big(\frac{6U_{c,d}}{L_{c,d}}\Big)$. In case 2, the relationship simplifies to $\alpha\leq\ln\Big(\frac{6U_{c,d}}{L_{c,d}}\Big)$. In case 3, the payment and remuneration fully cancel each other. As such, the Generalized Differential Allocation-Payment Relationship holds when choosing $\alpha = \alpha_{c,d}= \ln{\Big(\frac{6 U_{c,d}}{L_{c,d}}\Big)}$. Because \eqref{eqn:gendiffalloc} holds for the charging/discharging pricing functions, cost functions, and Fenchel conjugates, the remainder of the proof follows from Lemmas \ref{Background 2}, \ref{Background 3}, and \ref{Background 1}.\hfill$\square$

\vspace{10ex}

\begin{IEEEbiography}[{\includegraphics[width=1in,height=1.21in,clip,keepaspectratio]{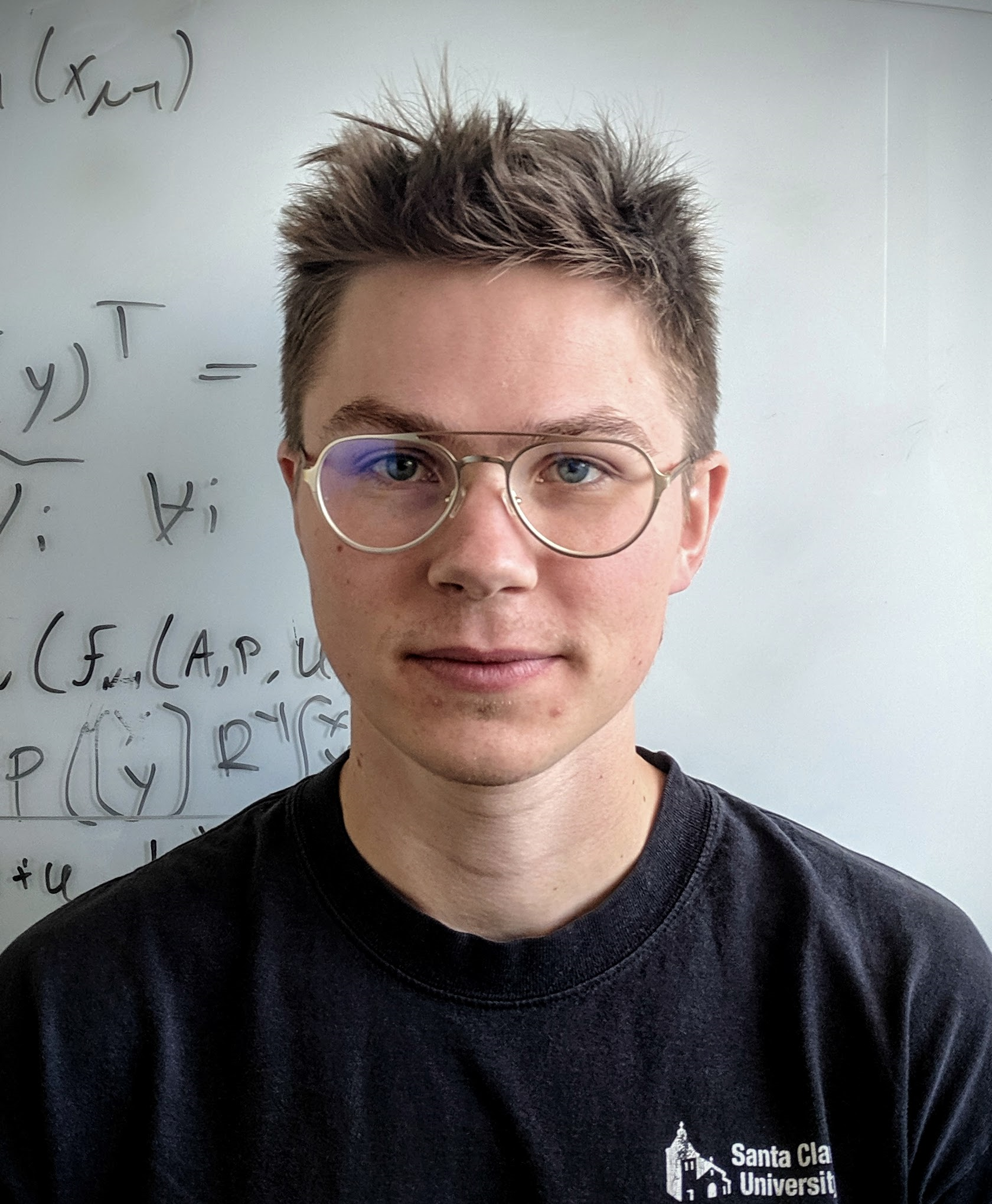}}]{NATHANIEL TUCKER}is a Ph.D. candidate in Electrical and Computer Engineering at the University of California, Santa Barbara. He received the B.Sc. degree in Electrical Engineering and Computer Science as well as the M.Sc. degree in Electrical Engineering from Santa Clara University in 2016 and 2017, respectively. His research interests include optimization and reinforcement learning for the design, control, and analysis of smart infrastructure systems such as the power grid and transportation systems.
\end{IEEEbiography}
\begin{IEEEbiography}[{\includegraphics[width=1in,height=1.21in,clip,keepaspectratio]{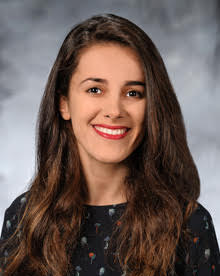}}]{MAHNOOSH ALIZADEH}is an assistant professor of Electrical and Computer Engineering at the University of California Santa Barbara. Dr. Alizadeh received the B.Sc. degree in Electrical Engineering  from Sharif University of Technology in 2009 and the M.Sc. and Ph.D. degrees from the University of California Davis in 2013 and 2014 respectively, both in Electrical and Computer Engineering. From 2014 to 2016, she was a postdoctoral scholar at Stanford University. Her research interests are focused on designing scalable control and market mechanisms for enabling sustainability and resiliency in societal infrastructures, with a particular focus on demand response and electric transportation systems. Dr. Alizadeh is a recipient of the NSF CAREER award. Dr. Alizadeh has served as an associate editor for the \textit{IEEE Transactions on Smart Grid} since 2020.
\end{IEEEbiography}
\end{document}